\documentclass[sigconf]{acmart}

\usepackage{xspace}

\usepackage{cleveref} 
\usepackage[inline]{enumitem} 

\usepackage{tcolorbox}  
\tcbuselibrary{listingsutf8}
\usepackage{subcaption}

\AtBeginDocument{%
  }

\copyrightyear{2025}
\acmYear{2025}
\setcopyright{rightsretained}
\acmConference[WebConf '25]{Proceedings of The Web Conference}{April 28 -- May 2, 2025}{Sydney, Australia}
\acmBooktitle{Proceedings of The Web Conference (WebConf '25), April 28 -- May 2, 2025, Sydney, Australia}
\acmDOI{XXXXXXX.XXXXXXX}

\acmConference[WebConf '25]{Proceedings of The Web Conference}{April 28 -- May 2, 2025}{Sydney, Australia}
\acmISBN{978-1-4503-XXXX-X/18/06}

\begin{document}

\title{Multi-stage Large Language Model Pipelines Can Outperform GPT-4o in Relevance Assessment}

\author{Julian A. Schnabel}
\email{julian.schnabel@hhu.de}
\orcid{0009-0001-5461-3254}
\affiliation{%
  \institution{Heinrich Heine University}
  \city{Düsseldorf}
  \country{Germany}
}

\author{Johanne R. Trippas}
\email{j.trippas@rmit.edu.au}
\orcid{0000-0002-7801-0239}
\affiliation{%
  \institution{RMIT University}
  \city{Melbourne}
  \country{Australia}
}

\author{Falk Scholer}
\email{falk.scholer@rmit.edu.au}
\orcid{0000-0001-9094-0810}
\affiliation{%
  \institution{RMIT University}
  \city{Melbourne}
  \country{Australia}
}

\author{Danula Hettiachchi}
\email{danula.hettiachchi@rmit.edu.au}
\orcid{0000-0003-3875-5727}
\affiliation{%
  \institution{RMIT University}
  \city{Melbourne}
  \country{Australia}
}


\begin{abstract}
The effectiveness of search systems is evaluated using relevance labels that indicate the usefulness of documents for specific queries and users. While obtaining these relevance labels from real users is ideal, scaling such data collection is challenging. Consequently, third-party annotators are employed, but their inconsistent accuracy demands costly auditing, training, and monitoring. We propose an LLM-based modular classification pipeline that divides the relevance assessment task into multiple stages, each utilising different prompts and models of varying sizes and capabilities. Applied to TREC Deep Learning (TREC-DL), one of our approaches showed an 18.4\% Krippendorff's $\alpha$ accuracy increase over OpenAI's GPT-4o mini while maintaining a cost of about 0.2 USD per million input tokens, offering a more efficient and scalable solution for relevance assessment. This approach beats the baseline performance of GPT-4o (5 USD). With a pipeline approach, even the accuracy of the GPT-4o flagship model, measured in $\alpha$, could be improved by 9.7\%. 
 
\end{abstract}

\ccsdesc[500]{Information systems~Information retrieval}

\keywords{Relevance Judgements, LLMs, Evaluation}

\maketitle

\section{Introduction}
\label{sec:intro}

Accurate document relevance labels are essential for training and evaluating retrieval systems, as they determine the effectiveness of search results in meeting user needs. However, labelling documents is a time-consuming and costly task~\cite{craswell2023overview, roitero2022preferences}. Currently, trained human assessors or crowd workers are employed to evaluate query-document pairs, but these approaches are often resource-intensive and prone to biases.
Recent research has proposed using LLMs for relevance assessment to reduce the dependence on time-consuming and costly human assessments while improving accuracy and alignment~\cite{liu2024information}.
However, employing large complex LLMs, like the flagship GPT-4o, would still incur high costs. 
We propose a modular classification pipeline, which has the potential to reduce labelling costs further while achieving similar accuracies compared to costly LLMs. 
Our pipeline approach is divided into two main steps; the LLM performs a binary classification followed by a more detailed three-level relevance labelling. This structured approach streamlines the labelling process by combining an initial relevance filter with granular classification. We demonstrate that this pipeline delivers comparable labelling accuracy to state-of-the-art LLMs while significantly lowering costs, offering a scalable and efficient solution for high-quality data annotation.

\begin{figure}[H]
    \centering
    \includegraphics[width=0.52\linewidth]{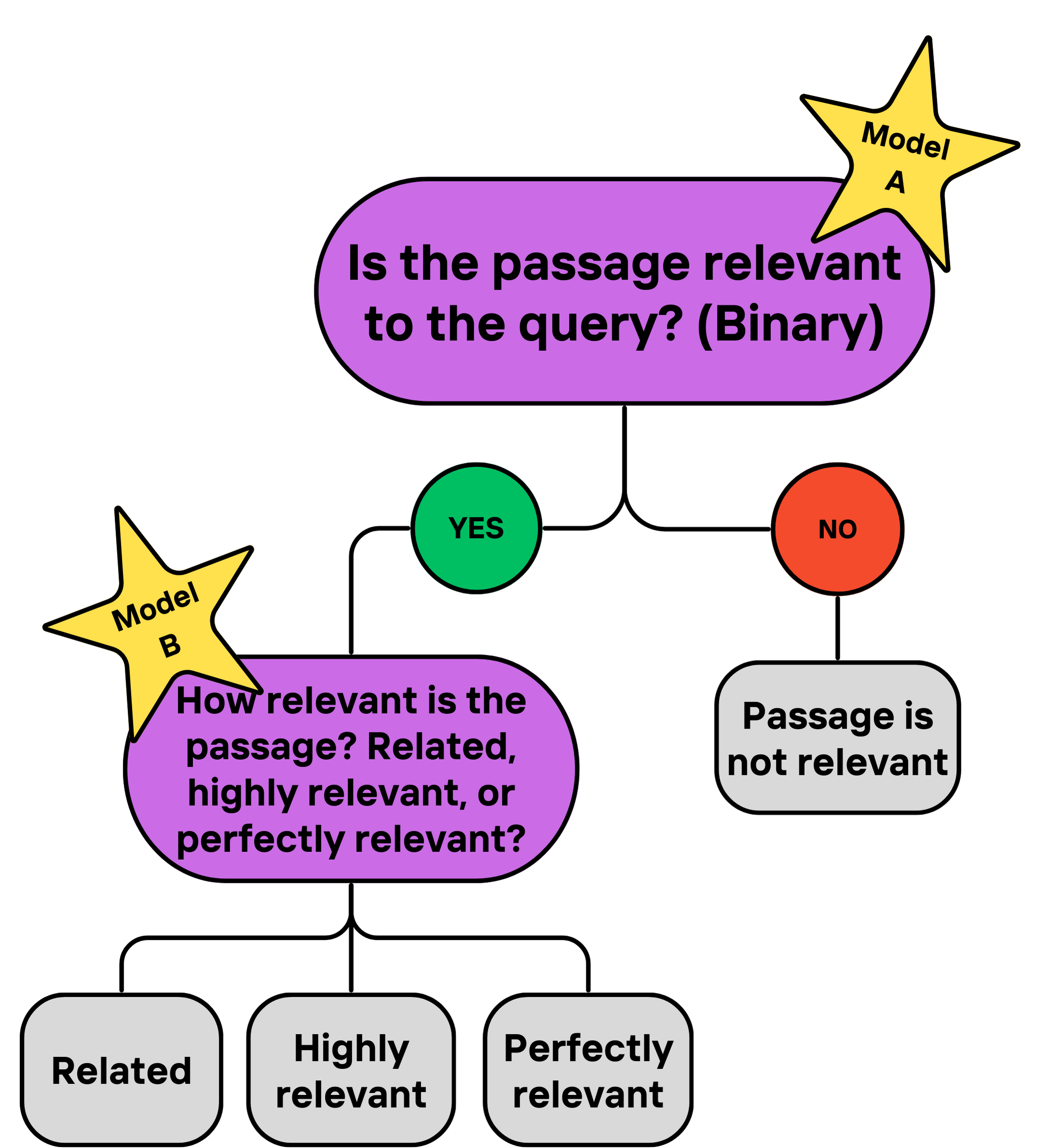}
    \caption{Visual overview of the pipeline approach where different models can judge at different stages of the relevance judgement labelling.}
    \label{fig:pipeline_vis}
\end{figure}

\section{Related Work}
\label{sec:relatedwork}
Several works have demonstrated the feasibility of using LLMs for relevance judgement. \citet{thomas2024large} explained the labelling process and prompts for using LLMs to judge relevance at Bing. \citet{upadhyay2024umbrela} provided an open-source reproduction of the Bing Relevance assessor~\cite{thomas2024large} using GPT-4o. They provided the prompt and baseline accuracy score we used to validate our methods.
\citet{Alaofi2024LLMs} presented a comprehensive performance overview of different models, both commercial and open-source, and also tested different prompting techniques. Their findings about the impact of adversarial prompt- and keyword injection on relevance judgement suggest the value of using a complex model at the last stages of a labelling pipeline. \citet{zendel2024enhancing} conducted experiments with batch processing and labelling, suggesting that batch processing could further reduce the cost of our proposed pipeline.
\section{Methodology}
\label{sec:methodology}

\subsection{Dataset and Evaluation}
\label{subsec:Deep Learning TREC Datasets}
We use the TREC 2023 Deep Learning Track (TREC-DL 23)~\cite{craswell2023overview}, similar to~\citet{upadhyay2024umbrela}. We use human relevance labels (i.e., \textit{Qrels}) provided for queries and assess them using our proposed systems. These relevance labels, commonly called ``gold labels'', are the benchmark for evaluating our systems. The original judgments were made by NIST assessors, who, given a query, assigned relevance scores to passages based on the following scale~\cite{craswell2020overview}:
\begin{enumerate}\addtocounter{enumi}{-1}
    \item \textbf{Irrelevant}: The passage has nothing to do with the query.
    \item \textbf{Related}: The passage seems related to the query but does not answer it.
    \item \textbf{Highly relevant}: The passage has some answer for the query, but the answer may be a bit unclear, or hidden amongst extraneous information.
    \item \textbf{Perfectly relevant}: The passage is dedicated to the query and contains the exact answer.
\end{enumerate}
Although these relevance labels are considered authoritative, assessors do not have the actual information need and must infer the user's intent when assigning relevance. This reliance on subjective judgment can introduce biases or inconsistencies, which may affect the accuracy of the gold labels in reflecting user relevance. In the TREC-DL 23 dataset, the label distribution is as follows: Label 0 ($13,866$), Label 1 ($4,372$), Label 2 ($2,259$) and Label 3 ($1,830$).

\paragraph{Models and Metrics} 
We used five models: 
\textit{GPT-4o} and \textit{GPT-4o mini} via Microsoft Azure, and \textit{Llama 3.1 70B instruct}, \textit{Llama 3.1 305B} and \textit{Claude 3.5 Sonnet} via OpenRouter\footnote{\url{https://openrouter.ai/}}. 
We did not perform any fine-tuning and used all the standard parameters. 
Only \textit{GPT-4o} and \textit{GPT-4o mini} were used to analyse the proposed pipeline approaches. The remaining models were only used for cost and accuracy comparison. When referencing binary accuracy, we use Cohen's $\kappa$; when referencing accuracy on a nominal scale, we use Krippendorff's $\alpha$.

\subsection{Reproducing Existing Baselines}
\label{subsec:Reproducing Existing Baselines}

\subsubsection{Zero-shot baseline from UMBRELA~\cite{upadhyay2024umbrela}}
\label{subsubsec:Zero-shot baseline from UMBRELA}
For our baseline, we use a zero-shot prompting technique with the \textit{description}, \textit{narrative}, and \textit{aspects} (DNA) method~\cite{thomas2024large} with the GPT-4o model. 
This DNA prompt is divided into three structured sections where \textit{description} and \textit{narrative} clarify the user query and the passage the LLM needs to label, helping establish context, and \textit{aspects} provides a step-by-step guide to structure the relevance labelling task into smaller, more manageable components, facilitating a more nuanced interpretation by the LLM.
To verify and reproduce the results presented by~\citet{upadhyay2024umbrela}, we used their exact prompt
(See Figure~\ref{baseline-prompt}) 
and run the same test on GPT-4o. We refer to this prompt as the ``\textit{Normal}'' prompt.

\subsubsection{Zero-shot baseline from UMBRELA run with GPT-4o mini}
\label{subsubsec:Zero-shot baseline from UMBRELA but with GPT-4omini}
Using the same dataset and the same \textit{Normal} prompt~\cite{upadhyay2024umbrela}, we reproduced the results with GPT-4o mini, OpenAI's budget LLM, with a cost per million input tokens of 0.15 USD instead of 5.00 USD for GPT-4o, the current flagship model.

\subsection{Relevance Judgement Pipeline Approaches}
\label{subsec:Relevance Judgement Pipeline Approaches}
We propose three novel relevance judgement pipeline approaches, categorised by single-stage vs. multi-stage and single-model vs. multi-model, as summarised in \Cref{tab:Proposed approaches}.
UMBRELA represents the single-model, single-stage method (i.e., baseline), while our proposed method uses a multi-model, single-stage approach. 
We also test a multi-stage (i.e., starting with a binary decision and refining to three relevance levels) approach for both single and multi-model single-model methods.

\begin{table}[htb]
\centering
\caption{Relevance assessment methods, categorised by single-stage vs. multi-stage and single-model vs. multi-model.}
\label{tab:Proposed approaches}
\begin{tabular}{lll}
\toprule
 & \textbf{Single-stage} & \textbf{Multi-stage} \\ \hline
\textbf{Single-model} & UMBRELA (\Cref{subsec:Reproducing Existing Baselines}) & \Cref{subsubsec:Single-model 2-stage judging from binary decision to three relevance levels} \\
\textbf{Multi-model} & \Cref{subsubsec:Mixture of Relevance Assessors} & \Cref{subsubsec:Multi-model 2-stage judging from binary decision to three relevance levels} \\ \bottomrule
\end{tabular}
\end{table}

\subsubsection{Multi-model Single-stage: Same prompt - different assessors}
\label{subsubsec:Mixture of Relevance Assessors}
In the Multi-model Single-stage approach, the \textit{Normal} prompt is used for two classification stages, each stage with a different model. 
First, we use the \textit{Normal} prompt to classify all Qrels. All Qrels deemed irrelevant (i.e., score 0) are excluded from further classification.
Next, we use the \textit{Normal} prompt again, but only the relevant Qrels are judged again.
This adjustment allows the second assessor (i.e., the second LLM) to classify a document that passed the initial irrelevance filter as irrelevant. \citet{Alaofi2024LLMs}~demonstrated that LLMs can be misled into labelling documents as relevant, a vulnerability particularly evident in smaller models. Assigning the second assessor -- typically the larger and more robust model -- the ability to override classifications made by smaller models could potentially mitigate misclassification caused by prompt manipulation or injection. However, this hypothesis remains untested, as~\citet{Alaofi2024LLMs} did not conduct additional experiments to validate it.

\subsubsection{Single-model Multi-stage Judging (from Binary to Three Relevance Levels)}
\label{subsubsec:Single-model 2-stage judging from binary decision to three relevance levels}

This approach uses one LLM for a two-stage relevance evaluation framework.
The first stage involves a \textit{binary classification} to determine if a passage is relevant to a given query; see Figure~\ref{modified-binary-prompt}.
If deemed relevant, the passage is classified into one of three relevance levels (i.e., related (1), highly relevant (2), perfectly relevant (3), see~\Cref{subsec:Deep Learning TREC Datasets}), enhancing the assessment's precision and granularity, see Figure~\ref{modified-3-scale-prompt}.
This structured approach, guided by the \textit{Normal} DNA prompt, ensures that the model provides a clear initial decision followed by a detailed categorisation of relevant passages.

\subsubsection{Multi-model Multi-stage Judging from Binary to Three Relevance Levels}
\label{subsubsec:Multi-model 2-stage judging from binary decision to three relevance levels}

Next, we use the approach explained in the previous section, but instead of using one model, different models are used for each of the stages (i.e., the binary and then three relevance level judgments).
Thus, this pipeline approach variant involves dividing the judgment process by the level of importance and \textit{assigning different models to each stage}. In practice, small and inexpensive models may perform well in making the binary decision of relevance versus irrelevance. Still, they may need assistance with the more complex task of distinguishing between documents rated as different relevance levels (i.e., highly relevant vs perfectly relevant). This approach has the potential to be cost-effective.

\section{Results}
\label{sec:results}

\Cref{tab:accuracy-different-combinations} summarises the evaluation outcomes of the baselines and all proposed pipelines. We evaluate two models (GPT-4o and 4o-mini) under both homogeneous (4o--4o, mini--mini) and heterogeneous (mini--4o, 4o--mini) pairings, alongside two prompt types: Binary--Relevant and Binary--Normal.

\begin{table}[h]
    \small	
    \captionsetup{skip=5pt} 
    \centering
    \caption{Accuracy for different GPT model/prompt combinations on TREC-DL23. Cost in USD per million input tokens.} 
    \begin{tabular}{cc|cc|c|cc|l} 
        \toprule
          \multicolumn{2}{c}{\textbf{Model}} & \multicolumn{2}{c}{\textbf{Prompt}}& \textbf{Binary} & \multicolumn{2}{c}{\textbf{4-scale}} &\textbf{Cost}\\
          1&2&   1&2&  $\kappa$ &  $\kappa$ & $\alpha$  &USD\\
          \bottomrule
 4o& -& Normal& -& 0.453& \textbf{0.296}&0.408 &5.00\\
 mini& -& Normal& -& 0.400& 0.254&0.359 &\textbf{0.15}\\ 
         \midrule
          mini&mini&   Binary
&Relevant
&  0.437
&  0.284& 0.422
 &0.21\\ 
          4o&4o&   Binary
&Relevant
&  0.428
&  0.280& 0.450
 &6.57\\ 
          mini&4o&   Binary
&Relevant
&  0.437&  0.286& 0.432 &2.05\\ 
          4o&mini&   Binary
&Relevant
&  0.428&  0.279& 0.443 &5.05\\ 
  mini&mini&   Binary &Normal
&  0.439&  0.281& 0.425 &0.21\\ 
         
 4o& 4o& Binary& Normal& 0.429& 0.280&\textbf{0.452} &6.57\\
 mini& 4o& Binary& Normal& 0.450& 0.295&0.446 &2.05\\
 4o& mini& Binary& Normal& 0.430& 0.276&0.445 &5.05\\
\midrule
 mini& 4o& Normal& Normal& 0.400& 0.260&0.367 &2.87\\
 4o& mini& Normal& Normal& \textbf{0.462}& 0.294&0.411 &5.05\\
 \bottomrule
    \end{tabular}
    \label{tab:accuracy-different-combinations}
    \vspace{-2ex}
\end{table}

\subsection{Single-model Single-stage Judging Results}
\label{subsec:Results}
\begin{figure}[H]
    \centering
    \includegraphics[width=1\linewidth]{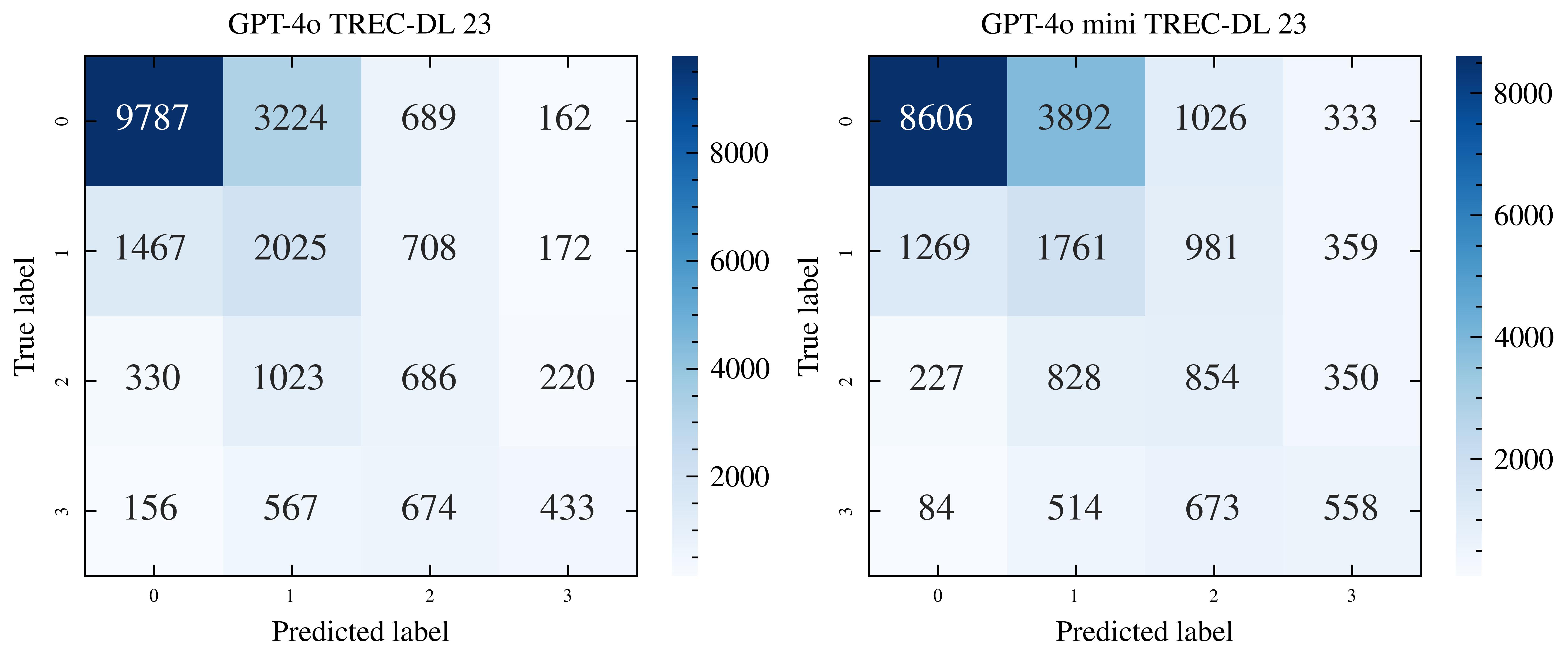}
    \caption{Reproduction of baseline (UMBRELA) with GPT-4o and GPT-4o mini.}
    \captionsetup{aboveskip=5pt}
    \vspace{-3mm}  
    \label{fig:4o-and-mini-umbrela-reproduction}
\end{figure}

The reproduced UMBRELA baseline~\cite{upadhyay2024umbrela} obtained similar accuracy scores, as shown in Figure~\ref{fig:4o-and-mini-umbrela-reproduction}. The misjudgement ``pattern'' is also similar to UMBRELA's. GPT-4o seems to be slightly over-optimistic.

GPT-4o still performs best among all tested stand-alone models. This aligns with the findings of~\citet{Alaofi2024LLMs}. However, GPT-4o mini's accuracy is, regarding the cost per input tokens being only 3\% of the cost of GPT-4o, satisfactory. As shown in Figure~\ref{cm-4o-vs-mini}, the agreement between GPT4o-mini and its larger variant, GPT-4o, is high. Notably, the binary accuracy given in~\Cref{tab:accuracy-different-combinations} increased for GPT-4o mini when using the custom binary prompt, whereas the accuracy for GPT-4o was reduced.

\subsection{Multi-model Single-stage Judging Results}
\label{subsec:Multi-model 1-stage Judging Results}

Using the same \textit{Normal} prompt for both stages with different assessors (i.e., LLMs) produced the highest binary accuracy (See \Cref{tab:accuracy-different-combinations}) but yielded below-average results for four-scale $\kappa$.

\begin{figure*}[h]
    \centering
    \includegraphics[width=0.9\linewidth]{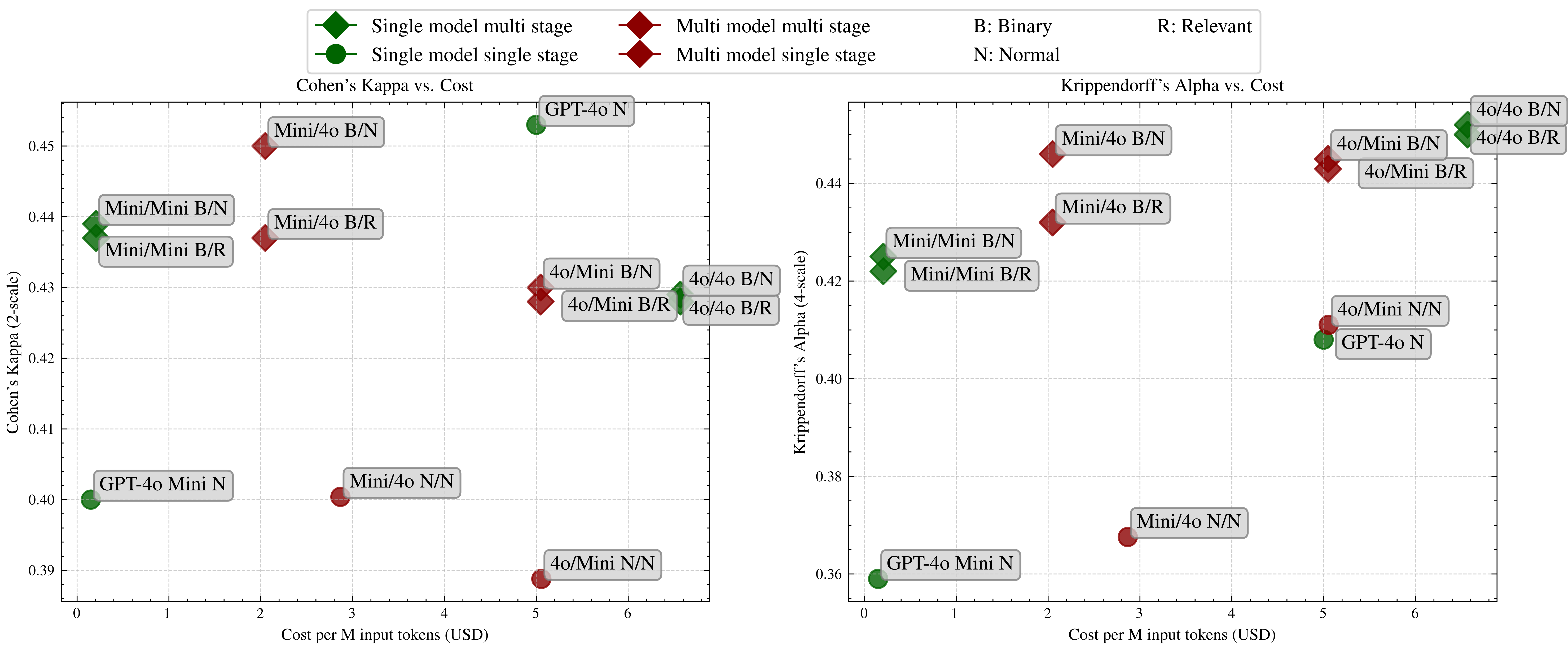}
    \caption{Cost vs Cohen's Kappa ($\kappa$) and Krippendorff's Alpha ($\alpha$).}
    \label{fig:Cost vs Cohen's Kappa}
\end{figure*}

\subsection{Multi-model Multi-stage Judging Results}
\label{subsec:Multi-model 2-stage Judging Results}

In the Multi-model Multi-stage approach using a modified binary prompt, the binary accuracy is identical to the binary accuracy of the respective Model 1. Four-scale Cohen's $\kappa$ is higher in every multi-model approach than single-model single-stage with GPT-4o. However, the (4-scale) $\kappa$-Score for GPT-4o could not be exceeded (See \Cref{tab:accuracy-different-combinations}). It is important to note that the GPT-4o mini/GPT-4o, binary/normal prompt combination reaches almost the same level of accuracy while significantly reducing the cost (See \Cref{fig:Cost vs Cohen's Kappa}).

Krippendorff's $\alpha$-score weights misjudgements on the difference to the actual score. Every Model/Prompt combination outperformed the baseline single model results. The highest $\alpha$-score for Multi-model Multi-stage was generated by utilizing the binary prompt with GPT-4o mini and the normal 4-scale prompt with GPT-4o as the relevance classifier. Similar accuracy was achieved by reversed model roles.

\subsection{Single-model Multi-stage Judging Results}
\label{subsec:Single-model 2-stage Judging Results}
\begin{figure}[htbp]
    \centering
    \begin{minipage}[t]{0.48\linewidth} 
        \vspace*{0pt} 
        \centering
        \includegraphics[width=\linewidth, keepaspectratio=true, height=5cm]{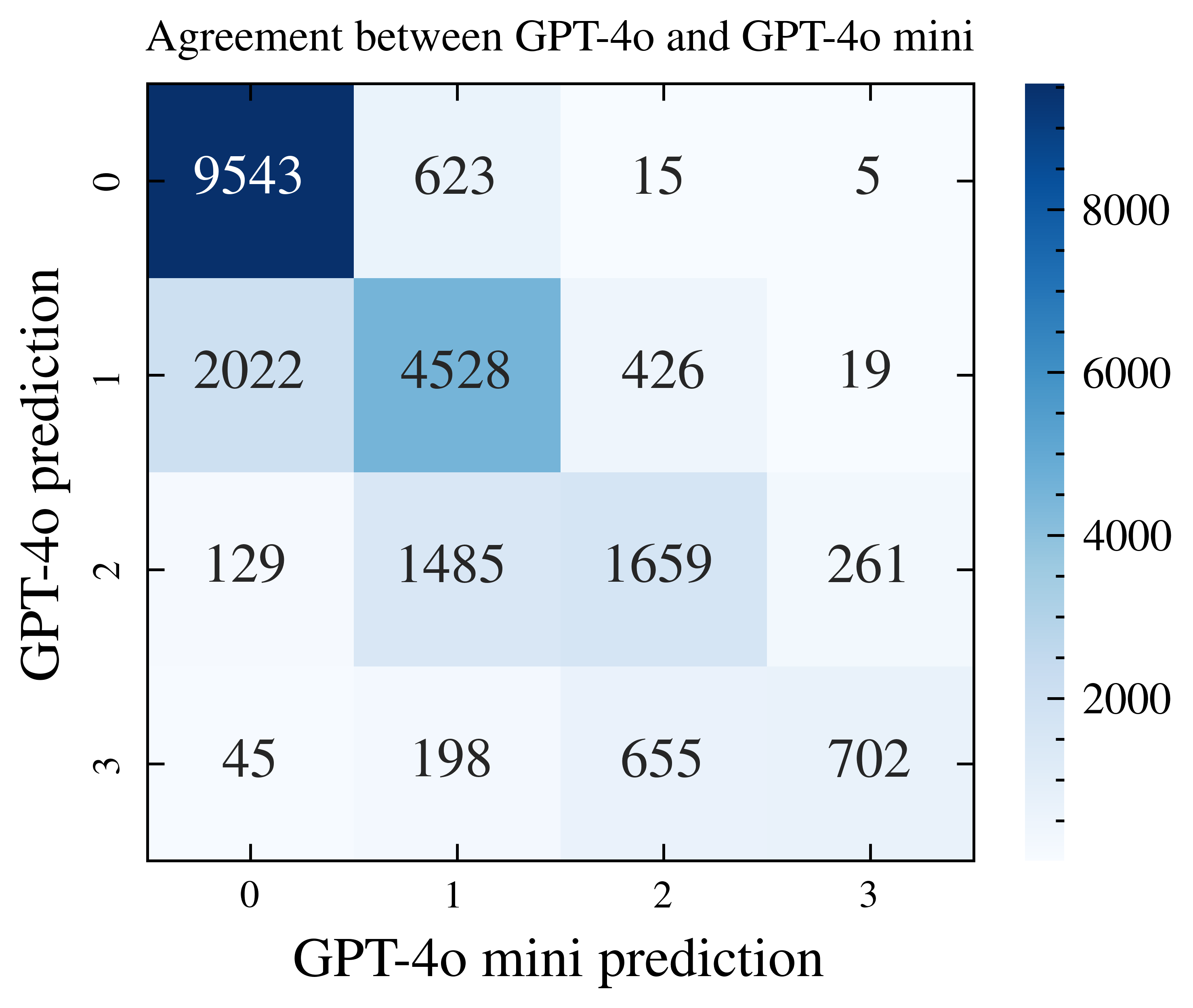}
        \caption{GPT-4o prediction vs GPT-4o mini predictions on TREC-DL 23.}
        \label{cm-4o-vs-mini}
    \end{minipage}%
    \hspace{\fill} 
    \begin{minipage}[t]{0.48\linewidth} 
        \vspace*{0pt} 
        \centering
        \includegraphics[width=\linewidth, keepaspectratio=true, height=5cm]{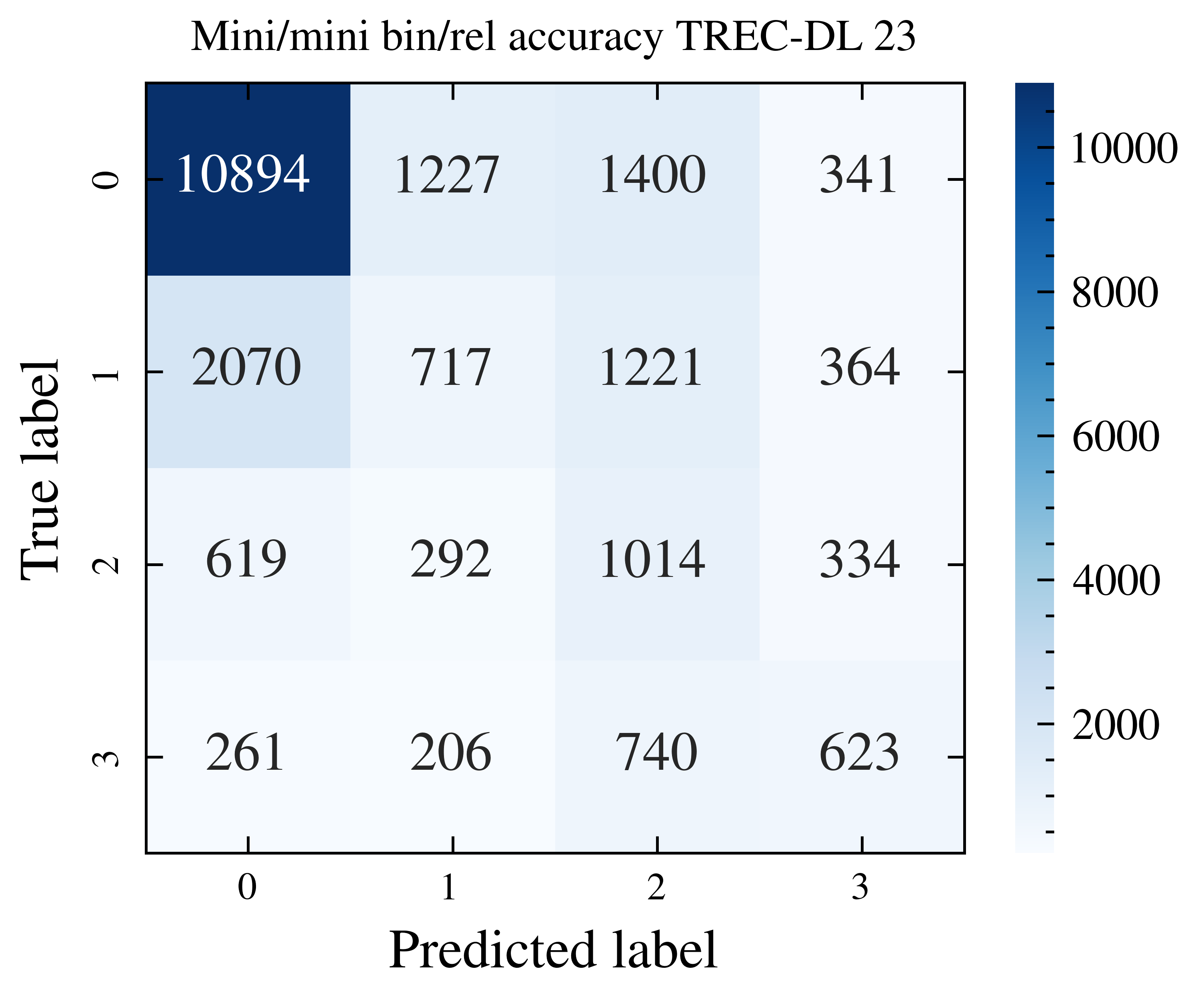}
        \caption{GPT-4o mini / GPT-4o mini Bin/R predictions on TREC-DL 23.}
        \label{fig:mireas-mini-mini-confusion-matrix}
    \end{minipage}
\end{figure}

Krippendorff's $\alpha$ increased for every version of the two-stage judgement for both GPT-4o and GPT-4o mini. Approaches, where both models were GPT-4o, performed slightly better than the small model pipelines.
However, using GPT-4o mini for both stages and the binary and normal prompt results in a 0.425 or 0.422 $\alpha$ when using the modified relevance prompt (Confusion matrix in \Cref{fig:mireas-mini-mini-confusion-matrix}). This exceeded the GPT-4o mini accuracy and even slightly outperformed the GPT-4o stand-alone. As shown in \Cref{fig:Cost vs Cohen's Kappa}, these results highlight that our pipeline approach can gain accuracy while significantly reducing costs.

\subsection{Cost Effectiveness}
\label{subsec:Cost Effectiveness}

For the cost calculation, we only consider the cost per million input tokens because only one character is generated as output. Thus, the output cost is negligible but would scale similarly to the input token cost. The following formula was used for the pipeline approach cost calculation, and the comparison of cost vs accuracy is shown in~\Cref{fig:Cost vs Cohen's Kappa}. 
$$
\text{Cost}= \text{cost}_{\text{M1}}+ \text{cost}_{\text{M2}}\cdot (1-\text{rate}_{\text{M1:} 0})
$$

\section{Conclusions} 
\label{sec:conclusion}
Of the proposed pipelines, except the Multi-model Single-stage approach, all combinations increase Krippendorff's $\alpha$ compared to the baseline. The largest increase in accuracy was achieved when using GPT-4o mini. Most notably, the Single-model Multi-stage approach achieved high accuracy while maintaining an extremely low cost. This accuracy could not be achieved just using GPT-4o, the flagship model, although it is more than 20 times more expensive. Regarding cost, the affordability of GPT-4o mini leaves room for more elaborate pipelines, incorporating even more stages. Possible approaches could include a step for each relevance label or more specialised prompts after a ``pre-classification''. 

Using a specialised binary classification prompts increased accuracy for the smaller model. Especially in the multi-stage approaches, dividing the relevance judgement task into a binary relevance decision and a relevance classification was beneficial for overall accuracy. 
We note limitations in our work. For example, \citet{upadhyay2024umbrela} states that near duplicates are contained in each label category in TREC-DL23. Since filtering these out is a complex process, we did not filter duplicates. In addition, to enhance the generalisability of findings, the tests could be conducted on more datasets, such as TREC-DL from different years and broader ranges of tasks. In addition, even though we reported in \Cref{subsec:Deep Learning TREC Datasets} that we used additional models to OpenAI, none of the open-source models demonstrated competitive performance compared to OpenAI's solutions. For this reason, we chose not to include their results in our analysis.
Further research could improve our prompt and/or fine-tune models for an even higher irrelevance-detection rate. For example, the prompt could be optimised with techniques such as chain-of-thought~\cite{wei2022chain} or narratives~\cite{sadiri2024unveiling}.
The TREC-DL datasets are heavily zero-weighted (ca. 75\%). This fact raises the importance of ``spam-filtering'' in relevance judgment tasks. Given that the current state of knowledge is that larger models perform better, a small (and affordable) ``spam-filter'' will significantly reduce overall assessment costs.

\begin{acks}
This research is partially supported by the Australian Research Council (CE200100005) and RMIT AWS Cloud Supercomputing~Hub.
\end{acks}

\bibliographystyle{ACM-Reference-Format}
\bibliography{strings,local}

\bigskip
\newpage
\appendix
\section{Prompts}

\begin{minipage}[htb]{\textwidth}
\centering
\begin{minipage}[b]{.49\columnwidth}
\begin{figure}[H]
    \centering
\footnotesize
\begin{tcolorbox}[
    colback=gray!20,           
    colframe=gray!50!black,    
    sharp corners,             
    boxrule=0.5mm,             
    fontupper=\ttfamily,       
    width=\linewidth,          
    listing only,              
    listing options={
        basicstyle=\ttfamily\footnotesize, 
        breaklines=true                    
    } 
    ]
Given a query and a passage, you must provide a score on an
integer scale of 0 to 3 with the following meanings:\\
0 = represent that the passage has nothing to do with the query,\\
1 = represents that the passage seems related to the query but
does not answer it,\\
2 = represents that the passage has some answer for the query,
but the answer may be a bit unclear, or hidden amongst extraneous
information and\\
3 = represents that the passage is dedicated to the query and
contains the exact answer.\\
Important Instruction: Assign category 1 if the passage is
somewhat related to the topic but not completely, category 2 if
passage presents something very important related to the entire
topic but also has some extra information and category 3 if the
passage only and entirely refers to the topic. If none of the
above satisfies give it category 0.\\
Query: \{query\}\\
Passage: \{passage\}\\
Split this problem into steps:\\
Consider the underlying intent of the search.\\
Measure how well the content matches a likely intent of the query
(M).\\
Measure how trustworthy the passage is (T).\\
Consider the aspects above and the relative importance of each,
and decide on a final score (O). Final score must be an integer
value only.\\
Do not provide any code in result. Provide each score in the
format of: \#\#final score: score without providing any reasoning.
\end{tcolorbox}
\caption{Baseline UMBRELA prompt as used by~\citet{upadhyay2024umbrela}.}
\label{baseline-prompt}
\end{figure}
\end{minipage}
\hfill
\begin{minipage}[b]{.49\columnwidth}

\begin{figure}[H]
    \centering
\footnotesize
\begin{tcolorbox}[
    colback=gray!20,           
    colframe=gray!50!black,    
    sharp corners,             
    boxrule=0.5mm,             
    fontupper=\ttfamily,       
    width=\linewidth,          
    listing only,              
    listing options={
        basicstyle=\ttfamily\footnotesize, 
        breaklines=true                    
    }  
    ]
Given a query and a passage, you must provide a score on an
integer scale of 1 to 3 with the following meanings:\\
1 = represents that the passage seems related to the query but
does not answer it,\\
2 = represents that the passage has some answer for the query,
but the answer may be a bit unclear, or hidden amongst extraneous
information and\\
3 = represents that the passage is dedicated to the query and
contains the exact answer.\\
Important Instruction: Assign category 1 if the passage is
somewhat related to the topic but not completely, category 2 if
passage presents something very important related to the entire
topic but also has some extra information and category 3 if the
passage only and entirely refers to the topic.\\
Query: \{query\}\\
Passage: \{passage\}\\
Split this problem into steps:\\
Consider the underlying intent of the search.\\
Measure how well the content matches a likely intent of the query
(M).\\
Measure how trustworthy the passage is (T).\\
Consider the aspects above and the relative importance of each,
and decide on a final score (O). Final score must be an integer
value only.\\
Do not provide any code in result. Provide each score in the
format of: \#\#final score: score without providing any reasoning.
\end{tcolorbox}
\caption{Modified 3-scale classification prompt (Relevant).}
\label{modified-3-scale-prompt}
\end{figure}
\end{minipage}
\par\bigskip
\par\bigskip
\begin{minipage}[b]{.49\columnwidth}
\begin{figure}[H]
    \centering
\footnotesize
\begin{tcolorbox}[
    colback=gray!20,           
    colframe=gray!50!black,    
    sharp corners,             
    boxrule=0.5mm,             
    fontupper=\ttfamily,       
    width=\linewidth,          
    listing only,              
    listing options={
        basicstyle=\ttfamily\footnotesize, 
        breaklines=true                    
    }  
    ]
Given a query and a passage, you must provide a score on an integer scale of 0 to 1 with the following meanings:\\
0 = represent that the passage has nothing to do with the query,\\
1 = represents that the passage has something to do with the query.\\
Important Instruction: Assign category 1 if the passage is relevant to the topic. If it is not relevant to the topic, assign category 0.\\
Query: \{query\}\\
Passage: \{passage\}\\
Split this problem into steps:\\
Consider the underlying intent of the search.\\
Measure how well the content matches a likely intent of the query (M).\\
Measure how trustworthy the passage is (T).\\
Consider the aspects above and the relative importance of each, and decide on a final score (0). The final score must be an integer value only.\\
Do not provide any code in the result. Provide each score in the format of: \#\#final score: score without providing any reasoning.
\end{tcolorbox}
\caption{Modified binary classification prompt (Binary).}
\label{modified-binary-prompt}
\end{figure}
\end{minipage}
\hfill
\begin{minipage}[b]{\columnwidth}
\end{minipage}

\end{minipage}
\end{document}